# Bonding, structures, and band gap closure of hydrogen at high pressures

Alexander F. Goncharov[1], John S. Tse[2,3], Hui Wang[2,3], Jianjun Yang[2], Viktor V. Struzhkin[1], Ross T. Howie,[4] and Eugene Gregoryanz[4]

[1]Geophysical Laboratory, Carnegie Institution of Washington, 5251 Broad Branch Road, Washington, D.C. 20015, USA

[2]Department of Physics and Engineering Physics, University of Saskatchewan, Saskatoon, SK S7N 5E2, Canada

[3]State Key Lab of Superhard Materials, Jilin University, 130012, Changchun, P. R. China

[4]Centre for Science at Extreme Conditions and School of Physic and Astronomy, University of Edinburgh, Edinburgh, EH9 3JZ, United Kingdom

**We have studied dense hydrogen and deuterium experimentally up to 320 GPa and using *ab initio* molecular dynamic (MD) simulations up to 370 GPa between 250 and 300 K. Raman and optical absorption spectra show significant anharmonic and quantum effects in mixed atomic and molecular dense phase IV of hydrogen. In agreement with these observations, *ab initio* MD simulations near 300 K show extremely large atomic motions, which include molecular rotations, hopping and even pair fluctuations suggesting that phase IV may not have a well-defined crystalline structure. The structurally diverse layers (molecular and graphene-like) are strongly coupled thus opening an indirect band gap; moreover, at 300 GPa we find fast synchronized intralayer structural fluctuations. At 370 GPa the mixed structure collapses to form a metallic molecular *Cmca*-4 phase, which exhibit a new interstitial valence charge bonding scheme.**

Hydrogen at high densities is remarkably rich in phenomena as revealed through both dynamic[1] and static[2-5] experiments and theoretical calculations[6-9]. Phases I and II have distinct quantum properties related to molecular rotations and possess species with different parity of the rotational states (ortho-para distinction)[10]. In contrast, phase III was proposed to be orientationally ordered in a classical sense[4,6], similar to molecular phases of heavier diatomic molecules, *e.g.* nitrogen. Surprisingly, the structure of a recently discovered phase IV has been found to consist of two very distinct structural units, as manifested by the presence of the Raman intramolecular vibrations (vibrons) with frequencies substantially different from each other and from other known phases of hydrogen.[5,8,11]

Based on the structural search of the lowest enthalpy phases, which included zero point motion and full vibrational contributions, a number of candidate structures were proposed for phase IV[8,11,12]. The theoretical studies all agree on the presence of two distinct structural units—a "molecular" ($Br_2$- like) and "graphene-like" (G) layers, which repeat periodically in the A-B-A-B sequence. However, the symmetry of the graphene-like layer, which consists of elongated weakly bonded $H_2$ molecules, is a matter of debate. It has been pointed out[11,12] that the originally proposed *Pbcn* structure[8], which contains 3 kinds of hexagonal rings, is dynamically unstable as the lowest frequency libron $(H_2)_3$ ring mode has an imaginary frequency. Two other very similar structures (*Pc*[11] and *Cc*[12]) have further been proposed as they have slightly lower enthalpy and show no imaginary frequencies.

Raman measurements of phase IV show a dramatic softening of the G-layer vibron mode with pressure[5] in a qualitative agreement with theoretical calculations[11] which reported that weakly bonded $H_2$ molecules elongate with pressure thus revealing a tendency of the G-layers to transform to truly graphene layers with symmetric hydrogen bonds. Thus, one can expect that phase IV would transform to the *Ibam* structure[11] at 350 GPa. However, the *Cmca*-4 structure[7,8] was found to be energetically favorable above 225 GPa[11,13], while metadynamics simulations[12] showed the transition to the same structure at 275 GPa. Both, the *Cmca*-4 and competing with it energetically *Cmca*-12 phases, are metallic[7,8,11,14] and their occurrence in the calculations have been invoked to explain the experimental observations of atomic metallic fluid hydrogen in Ref. 15. However, the findings of Ref. 15 are found to disagree with the optical data of Ref. 5.

In this paper, by combining the experimental data of Ref. 5 with *ab initio* molecular dynamical (MD) theoretical calculations we demonstrate fluxional characteristics of phase IV, which makes the description of the instantaneous structure of phase IV difficult in terms of the conventional space groups. This is demonstrated by experimental observations of significant quantum effects in molecular vibrations and librations in the G-layer, which are consistent with frequent molecular rotations and intermolecular atomic fluctuations. These findings are supported by MD calculations, which show very large atomic motions and intralayer $H_2$ re-arrangements at 250-300 K. MD simulations also demonstrate the transformation to the *Cmca*-4 structure at 370 GPa, where the structural diversity of the $Br_2$-like and G layers vanishes and large atomic motions cease.

Constant pressure-constant temperature (*N,P,T*) ensemble classical molecular dynamics calculations were performed with the pseudopotential (PS) plane wave method. The quality of the PS has been checked against all electron calculations to guarantee the results are valid within the pressure range studied here. A 1x1x2 *Pbcn*[8] supercell with 48 atoms was used as the initial structure. We have performed MD calculations for phase IV of hydrogen at 250 – 370 GPa at 250 - 300 K. Thermodynamic equilibrium was usually achieved in the first 6 ps. The length of the simulation varied from 20 ps at 250 and 370 GPa to 87 ps at 300 GPa. The maximum deviation throughout the duration of the MD calculations is less than 0.4 meV/atom in the internal energy and the pressure was within 2% of the target. The results of the simulations should be at least qualitatively correct since the ratio of the mean atomic distance $(V/N)^{1/3}$ and the thermal de Broglie wavelength is larger than 1 for the studied density range.[16] Quantum effect normally enhances delocalization of the hydrogen spatial distribution. Since the classical simulations were performed at relatively high temperature (*ca*. 300K), it is expected that the quantum zero-point motions will not fundamentally alter the hopping and molecular rearrangement processes.[6] *GW* calculations were performed with the code VASP[17], employing the PAW potential.[18] The band structure was constructed from interpolation of *GW* corrected Monkhorst-pact *k*-point set using Wannier functions.[19]

The detailed experimental procedures are described in Ref. 5. We have carried out an intensity calibration performed using a light source with a known spectral distribution to the Raman spectra measured (Fig. 1a) and applied a coupled oscillator model[20] to the intensity corrected Raman spectra. An interesting feature of the normalized Raman spectra is that they are dominated by the vibron modes. The model calculations show that the observed pressure induced relative change in intensity of some of the Raman modes with frequencies below 1300 cm$^{-1}$ (300, 800, and 1050 cm$^{-1}$ bands at 240 GPa) can be explained by their vibrational coupling with the lower frequency vibron mode from the G-layer, which softens with pressure abruptly (soft vibron mode). However, the 500 and 700 cm$^{-1}$ bands do not show any measurable intensity variation with pressure (Fig. 2). We interpret this diverse mode behavior as due to their different vector of the normal modes of vibrations (Fig. 3). The latter two modes correspond to the molecular translation and rotations in the Br$_2$ like layers, while the former ones, which show a substantial variation in intensity with pressure, correspond to similar motions in the G-layers. The 300 cm$^{-1}$ band is a unique G-layer mode corresponding to the librational motion of the hydrogen rings consisting of 3 $H_2$ dimers. Our mode assignment is based on theoretical calculations for $Pc$[11] and $Pbcn$[8] (Fig. 3) structures of hydrogen.

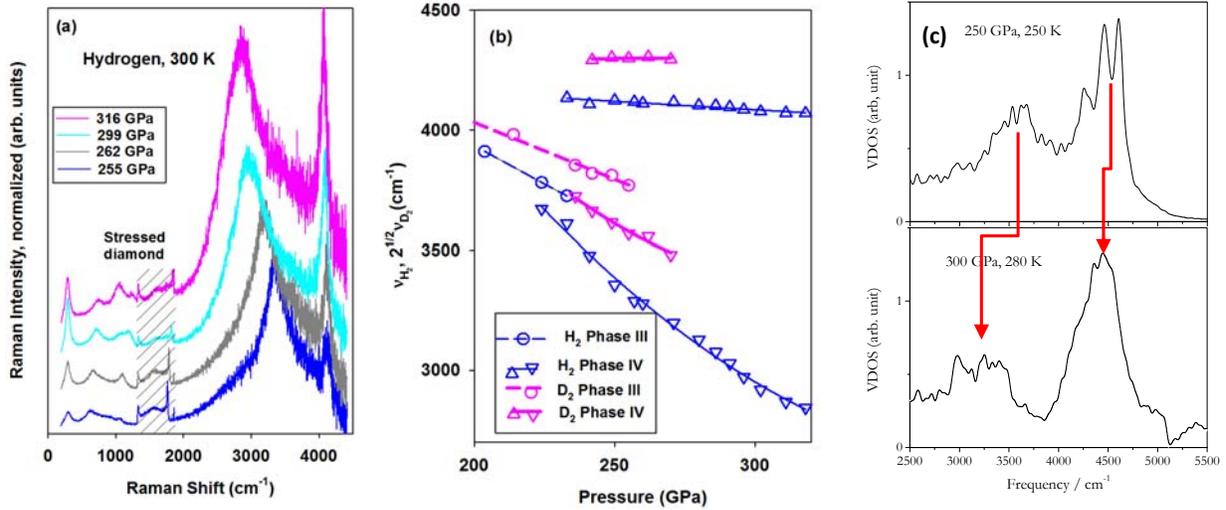

**Figure 1. Vibrational properties of $H_2$ and $D_2$ at 300 K up to 320 GPa. (a) Intensity normalized Raman spectra with pressure; the hatched area corresponds to Raman signal of stressed diamonds; (b) experimental Raman frequencies of the soft vibron modes in $H_2$ and $D_2$ with pressure; the frequencies of $D_2$ are multiplied by $\sqrt{2}$; discontinuities of the vibron frequencies at 230 GPa are due to the III-IV transition; there is a 10-15 GPa pressure range of phase coexistence; (c) vibrational spectra deduced from MD simulations. The vertical red lines show the behavior of two major Raman vibrons.**

The vibron frequencies show a distinct isotope effect for the vibron (Fig. 1b) and the lowest frequency libron mode (300 cm$^{-1}$), while the other modes show very small isotope dependencies (Figs. 4, 5). The isotope effect on the vibron and libron modes increases with pressure. For the vibron modes, the isotope effect is substantially larger for the soft mode. The linewidth of the soft vibron mode increases very steeply at 230-250 GPa and then remains almost pressure independent up to 320 GPa [Fig. 4(b)].

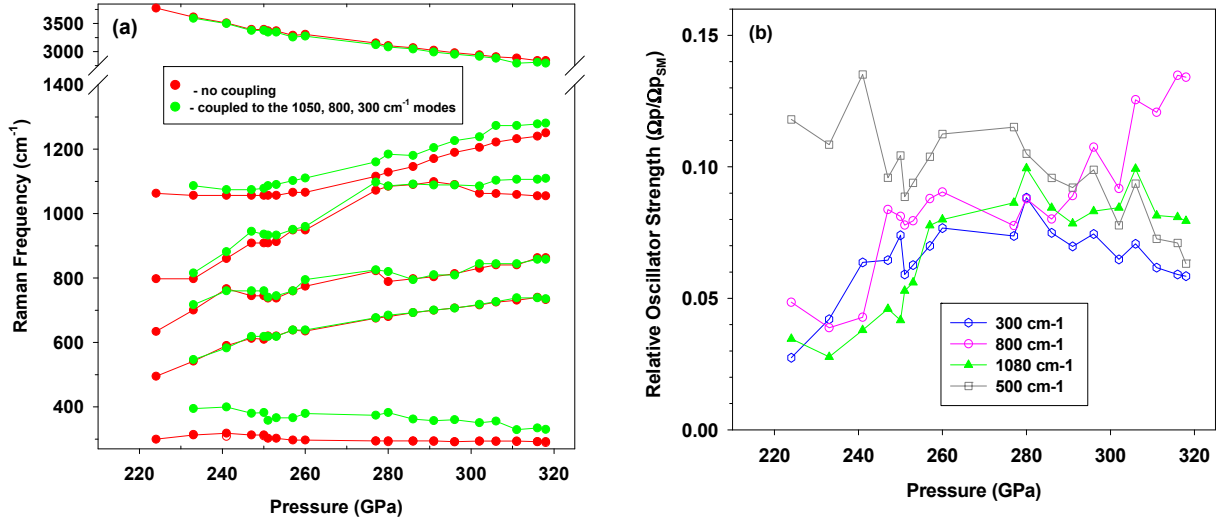

**Fig. 2. The results of the Raman modes analysis performed using a coupled oscillator model[20]. The panel (a) shows the positions of the experimentally measured Raman bands as a function of pressure determined by applying this model to the intensity corrected Raman spectra. The positions of the bare frequencies are given in two cases: (i) the 300, 800, and 1050 cm$^{-1}$ modes are coupled to the soft mode and the intensities of these modes are fully due to the coupling (green circles); (ii) there is no coupling (red circles), the positions of the bands and their intensities correspond to the measured ones. These cases correspond to two extremes; we expect that the reality is somewhere in between. The panel (b) shows the relative intensities of the four most intense low-frequency Raman bands with respect to that of the soft vibron mode. This illustrates the diverse behavior of the modes with respect to the vibrational coupling, which is expected to increase with pressure as the soft vibron mode and other modes become closer in frequencies.**

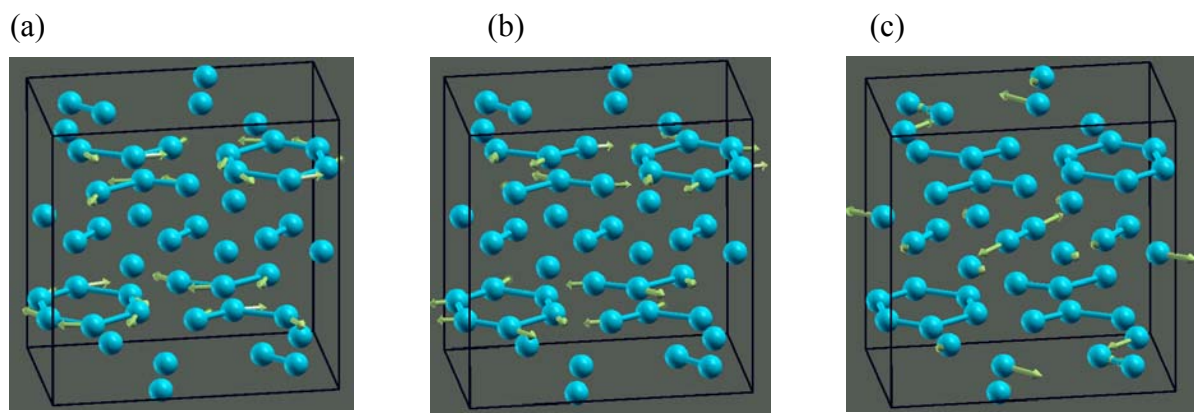

**Fig. 3. The atomic motions for the most prominent Raman modes of *Pbcn* hydrogen as the result of the DFT calculations. (a) low frequency libron mode; (b) soft vibron mode of the graphene-like layer; (c) vibron mode of the Br$_2$- like layer.**

A very large isotope effect and broadening (compared to other hydrogen phases) of the soft vibron mode from the G-layers is evidence for the dramatically increased anharmonicity of this mode. Using the Morse formalism for an underlying effective potential[21] we determined the pressure-dependent parameters of this potential using the experimental data for H$_2$ and D$_2$ [Fig.

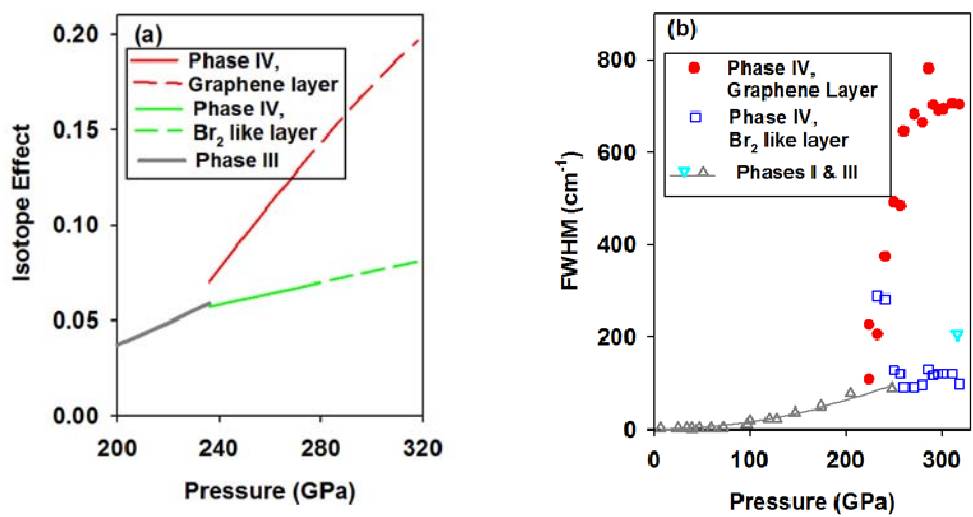

**Fig. 4. (a) The Raman isotope effect on the vibron frequency determined as** $\alpha = \log_{10}(\sqrt{2}\nu_{D_2}/\nu_{H_2})/\log_{10}(2)$ **; the dashed lines correspond to linear extrapolations of the D$_2$ data to higher pressures; (b) vibron linewidths of H$_2$ as a function of pressure. Data for phases I and III are from Refs. 22, 23 (gray triangles and cyan triangle, respectively).**

1(b)]. We find that at 320 GPa, the potential well becomes very shallow due to the lowering of the barrier between the potential wells; it can only accommodate only 2 vibrational levels for hydrogen (Fig. 6). The proximity of the first excited level to the barrier top splits this level as the tunneling between the wells becomes very frequent.[24] This provides a qualitative explanation for the anharmonic broadening effects [Fig. 4(b)]. We would like to stress that these effects have truly quantum characteristics, as much smaller peak broadening is observed for D$_2$ and there is a very large isotopic frequency difference for both the soft vibron and roton mode in the G-layer (Figs. 4-5). The vibrational properties of H$_2$ deduced from MD simulations (Fig. 1c) show substantial softening and broadening of the G-layer vibron at 3100-3600 cm$^{-1}$ between 250 and 300 GPa, which is in qualitative agreement with the experiment.

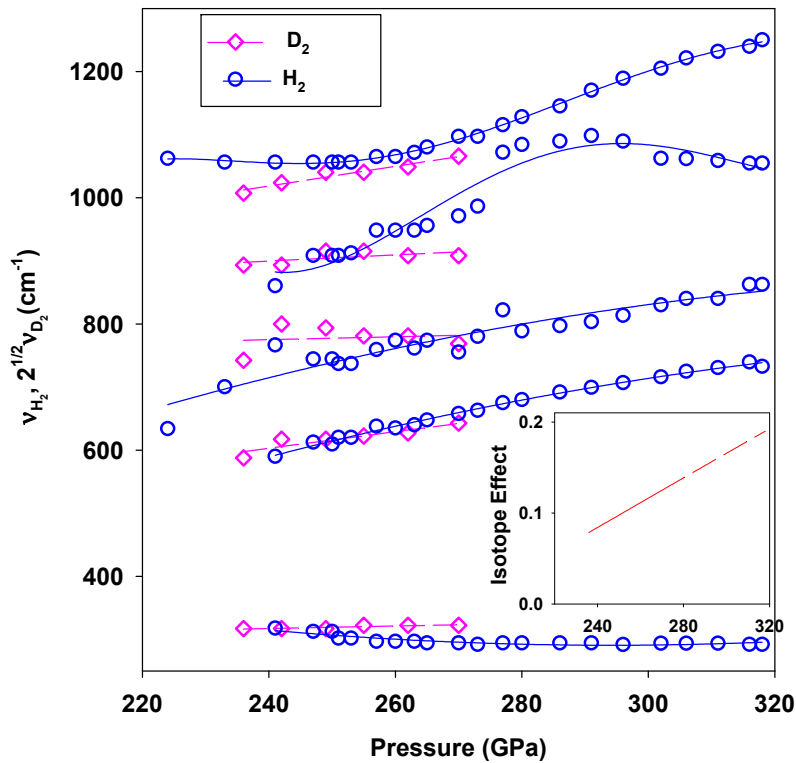

**Fig. 5. Raman frequencies of the lattice and librational modes of H$_2$ and D$_2$ with pressure; the Raman frequencies of D$_2$ are multiplied by $\sqrt{2}$. Inset: the Raman isotope effect on the lowest frequency libron frequency determined as $\alpha = \log_{10}(\sqrt{2}\nu_{D_2}/\nu_{H_2})/\log_{10}(2)$ ; the dashed lines correspond to linear extrapolations of the D$_2$ data to higher pressures.**

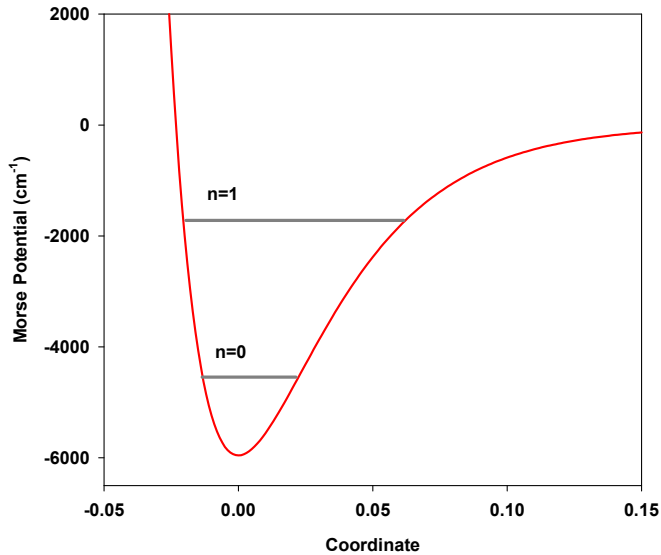

**Fig. 6. Effective intramolecular potential for dense hydrogen at 320 GPa determined from the Raman data for the soft vibron mode of H$_2$ and D$_2$ [Fig. 1(b)].[21] The data for D$_2$ are extrapolated using the best second order polynomial fit shown in Fig. 1(b). The potential is approximated by the Morse function and is plotted as a function of a dimensionless linear**

coordinate.[21] Horizontal lines represent the vibrational levels. Only two lowest energy vibrational levels correspond to the bound states at 320 GPa (solid lines).

Fig. 7 summarizes the results of the analysis of the MD trajectories for two dissimilar planes. At 250 GPa, the $Br_2$ like layers consist of a hexagonal lattice of "clouds" (Fig. 7, leftmost columns). These correspond to almost free rotation of $H_2$ molecules about the center of their bonds and maintaining the intramolecular bond lengths characteristic of "free" molecules (0.72 Å). In comparison, the G-layers can be represented as an almost equally populated honey-comb lattice, and one does not see any difference in dimensions or shape of the constituting hexagons, which are expected for the *Pbcn*[8] and *Pc*[11] structures. Further inspection include the color of dots (tracing the atoms) as the change in color would correspond to collaborative "hopping" of the $H_2$ molecules in the $Br_2$ layers and H atoms in the G-layers from one site to the other. At 250 GPa, this hopping is relatively infrequent for the $Br_2$ layers, while it seems more frequent for the G-layers with an average of ca. 1 jump every 4 ps. At 300 GPa (Fig. 7, central columns), the time average difference in distribution of the H atoms between $Br_2$ and G-layers apparently disappears; both show a honeycomb structure superimposed by the increased "clouds" due to molecular rotations. Closer analysis of the MD trajectories shows that this occurs due to a synchronized intralayer atomic fluctuations, which change the molecular-like $Br_2$ layers to the more atomic like G-layers and *vise versa*. Further details of the analysis will be published elsewhere. Based on extraordinary large atomic fluctuations observed in the MD runs at 250 and 300 GPa, we conclude that the structural presentation of phase IV in terms of an ideal crystal lattice *may not* be valid. The averaged over time crystal structure can be represented by the *Ibam* phase[8] with 8 atoms in the primitive unit cell.

Simulations at 370 GPa and 300 K (Fig. 7, rightmost columns) show a transition to another phase, which is very different from phase IV structurally and chemically (cf. *Ibam*). The amplitudes of the atomic motions are much smaller for this structure and all the crystallographic layers become structurally equivalent. This phase is metallic and it has *Cmca*-4 symmetry, which has been previously proposed.[7,8] The structure consists of crystallographically equivalent molecules with the bond length of 0.776 Å. The frequencies of the vibron in the *Cmca*-4 phase merge to a single set of modes at 3100 cm$^{-1}$ (Fig. 8). As will be discussed later, this is a consequence of weaker H-H bonds due to electron transfer from the bonding to the interstitial region.

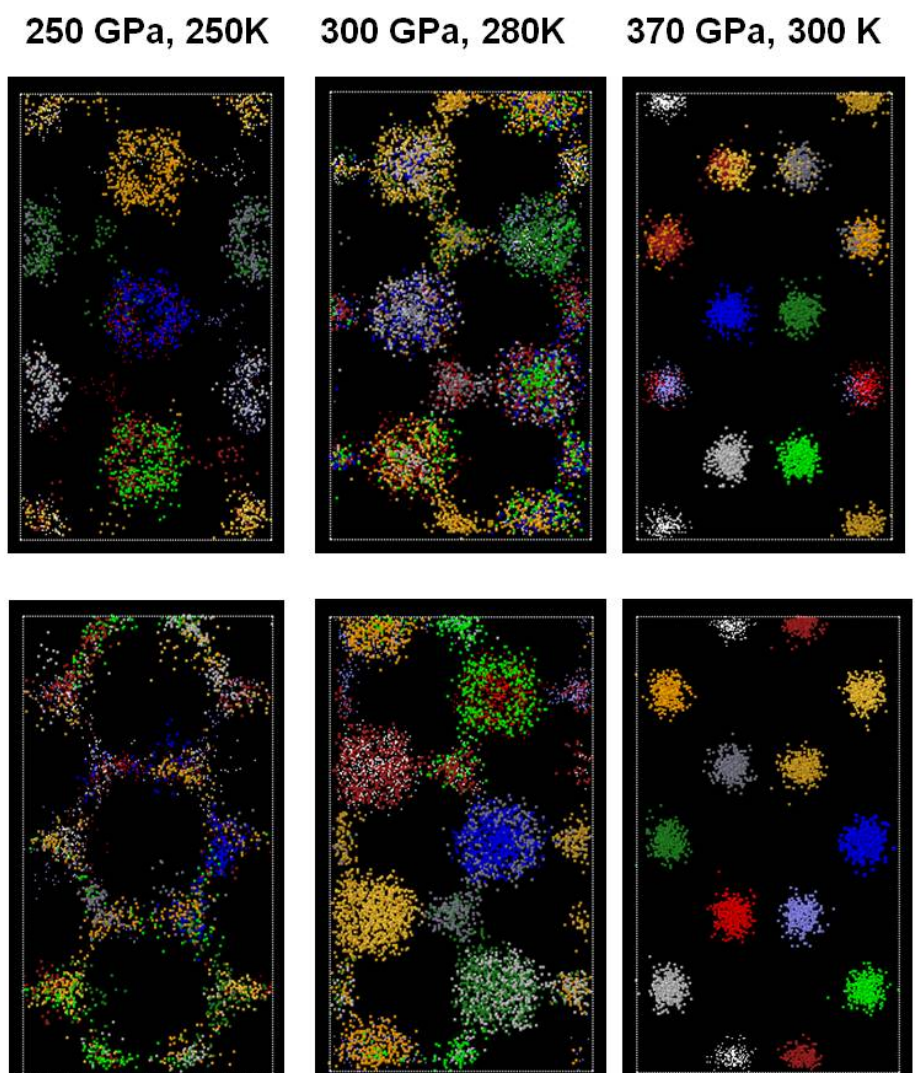

**Figure 7.** Molecular dynamics trajectories of three runs. The top row corresponds to the Br$_2$ like layer, while the bottom row to the G-layer. Different colors correspond to different atoms.

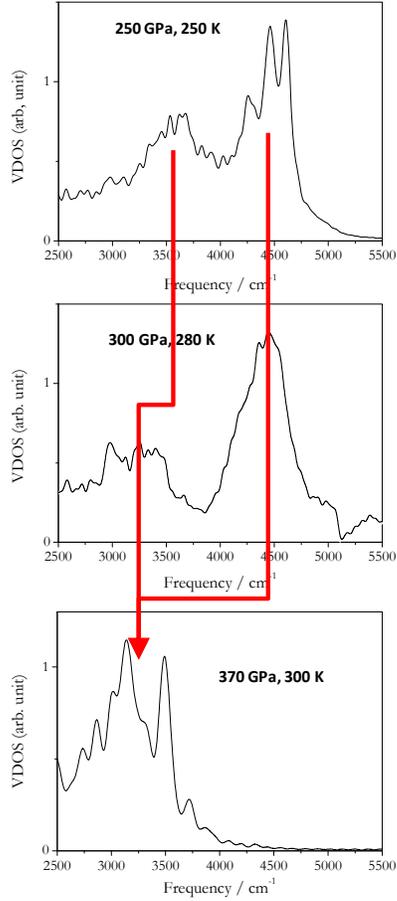

**Fig. 8.** The vibrational spectra deduced from MD simulations (the lines are guides to the eye showing the evolution of the H$_2$ vibrons with increasing pressure).

Now, we switch to the electronic properties. Optical absorption spectra [Fig. 9(a)] near the absorption edge demonstrate that the expression $\alpha \propto (h\nu)^2$ holds. This suggests that the absorption edge is determined by indirect optical transitions. This is in agreement with theoretical calculations, which clearly demonstrate that the band gap is indirect as the extrema of the conduction and valence bands are not in the Γ point (Fig. 10). Also, the experimental and theoretical (GW) pressure dependencies of the indirect band gap agree very well [Fig. 9(b)].

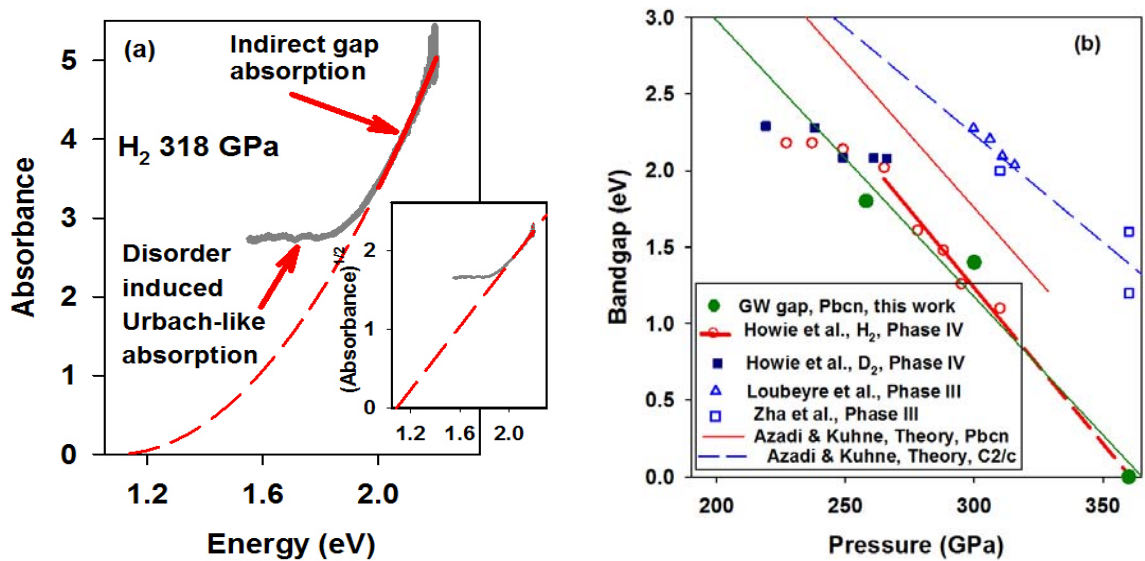

Figure 9. Band gap of phase IV of $H_2$. (a) experimental optical absorption spectra at 318 GPa. The inset shows that the square root of absorbance is linear with energy near the absorption edge, which is characteristic of indirect gap absorption. (b) experimental and theoretical band gaps for phases III and IV of $H_2$ and $D_2$. The data presented correspond to the following references: Howie et al.[5] reanalyzed here assuming that the band gap is indirect **resulting in lower values than in the original paper**, Azadi & Kuhne[25], Loubeyre et al.[23], and Zha et al.[26]

Theoretical calculations of the band gap using the hybrid DFT techniques[25] agree well with our data (Fig. 9), while those determined by standard DFT methods[11] show much lower band gaps (not shown). Also, our theoretical calculations do predict the bandgap closure at approximately the same pressure, at which the transition to the Cmca phase occurs in MD simulations. This may be purely coincidental as we have performed calculations for a limited number of pressure points. Moreover, the calculated band gaps are for the static *Pbcn* structures, while we believe that the structure of phase IV is fluxional. More investigations should be performed to find out whether this good agreement of the experimental and theoretical *GW* and hybrid DFT bandgaps is meaningful. The experimental spectra also show additional lower energy absorption (Fig. 9a), which we attribute to disorder induced processes similar to the Urbach absorption in disordered semiconductors. Similar effects have been observed previously in a disordered high-pressure phase of nitrogen[27]. Notably, the experimentally determined band gap of phase IV appears to be lower than that for phase III.

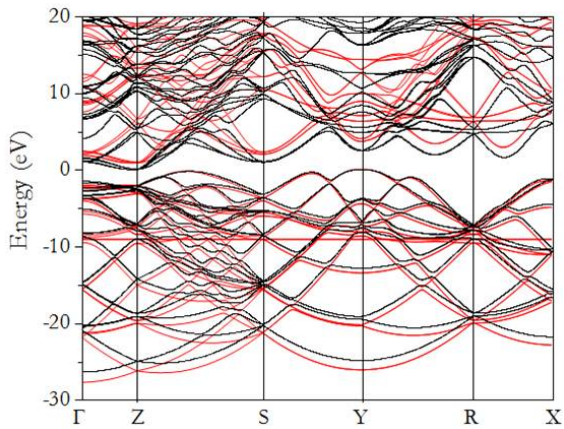

**Fig. 10. Electronic band structure of *Pbcn* $H_2$ at 300 GPa. Black and red lines are the results for GGA and *GW* methods, respectively.**

To investigate the nature of the chemical interactions on the electronic structure, calculations were performed on independent $Br_2$-like and G-layers in the *Pbcn* structure and compared with the full crystal (Figs. 11). It is surprising that both the $Br_2$ and G-layers are metallic! However, the interactions of these two distinct layers led to an opening of the band gap in the *Pbcn* structure. This finding can be explained by a charge transfer from the graphene layer to the $Br_2$ layer in the full structure. The valence band width was increased by almost 5 eV as a result of the interaction of the $Br_2$ and G-layers. This is accompanied by the opening of a small energy gap between the valence and conduction bands at 250 and 300 GPa.

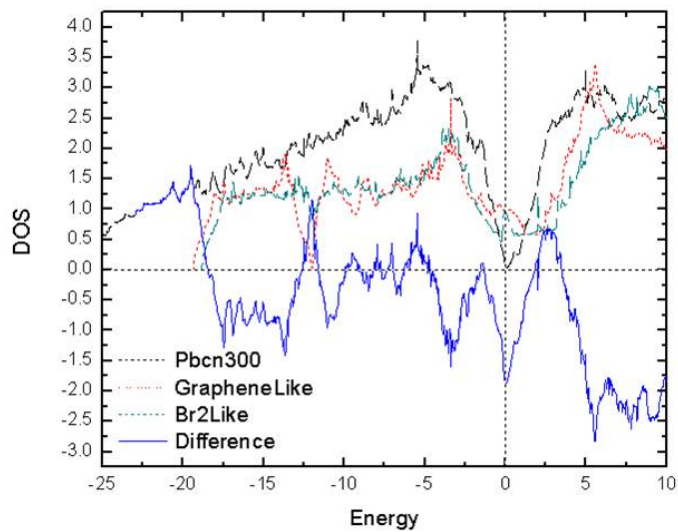

**Fig. 11. Full and partial density of electronic states of *Pbcn* hydrogen at 300 GPa.**

The electronic structures of both high-pressure high-temperature phases of hydrogen, phase IV and *Cmca*-4, are unusual for simple molecular solids. Plots of the difference in total charge density of the *Pbcn* structure from the constituent hydrogen atoms are shown in Fig. 12. Notably, the electron density is depleted in the atomic regions near the interstitial sites of the graphene layer. Therefore, there is a charge transfer from the G-layers to the $Br_2$ layer resulting in insulating behavior, which helps to stabilize the novel structure. Bader analysis of the electron density topology of the structure at 250 and 300 GPa (Figs. 13) shows the electrons in the graphene layer are not delocalized, and bond critical points between pairs of H atoms can be clearly defined. Thus, phase IV of $H_2$ remains molecular in nature. However, these molecules must be extremely short-lived judging from the results of MD simulation (Fig. 7), showing the extreme mobility of atoms in the G-layers. Moreover, at 300 GPa we even see the interlayer structural fluctuations (Fig. 7), which are very fast (<10 fs).

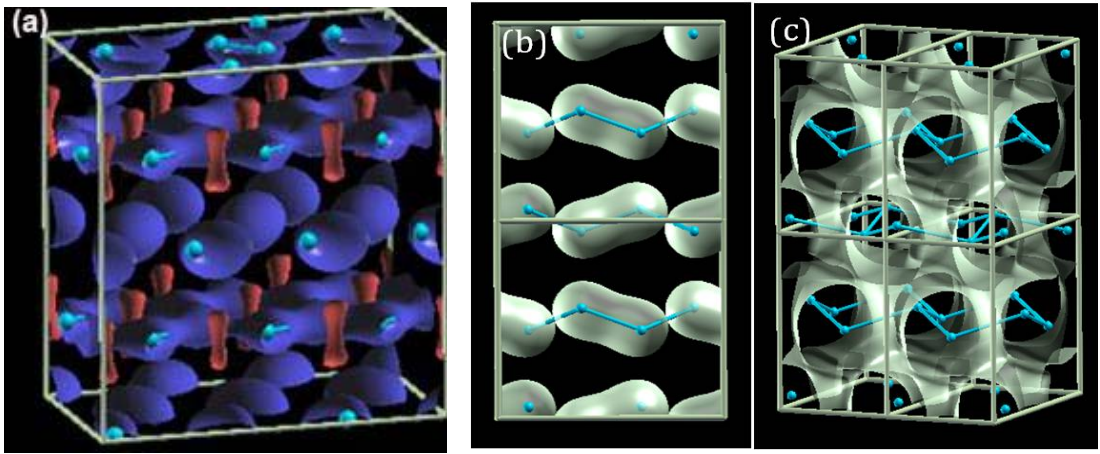

**Figure 12. Electron topology of high-pressure phases of $H_2$. (a) total electron density difference for *Pbcn* from the constituent hydrogen ($\Delta\rho$ = -0.176 e/Å$^3$). (b, c) Electron density iso-surfaces for the values of 1.0 e/Å$^3$ and 0.5 e/Å$^3$, respectively for the *Cmca*-4 structure.**

Finally, we analyze chemical bonding in the *Cmca*-4 structure, which becomes energetically more favorable than *Pbcn* above 320 GPa in our calculations; our MD simulations show that it forms at 370 GPa spontaneously. Electron density topological analysis[28] reveals that the structure of *Cmca*-4 is truly molecular but the electronic structure is metallic. The electronic density is partially pushed out of the intramolecular space to interstitials Fig. 12(b), thus resulting in decreased vibron frequency [Fig. 1(c)] with the rather short closest interatomic distance of 0.776 Å. The hypothetic high-pressure structure of $H_2$ (*Cmca*-4) demonstrates the presence of interstitial space "pockets" with an increased electronic density. Such phenomena have been previously reported under pressure in other light elements, for example in Li[29], but this is the first indication of that kind of behavior in hydrogen. Similar to the light alkalis, the interatomic potentials in hydrogen evolve with density to become very flat prior to this change in chemical bonding scheme (Fig. 6). Light elements in this regime tend to show structural diversity,[29-31] low-frequency vibrational modes[22,32], and low melting temperatures.[30,31] This is the consequence of proton zero point energy becoming the important energy scale because of a decline of the intramolecular bonding (*e.g.*, Refs. [22,31]).

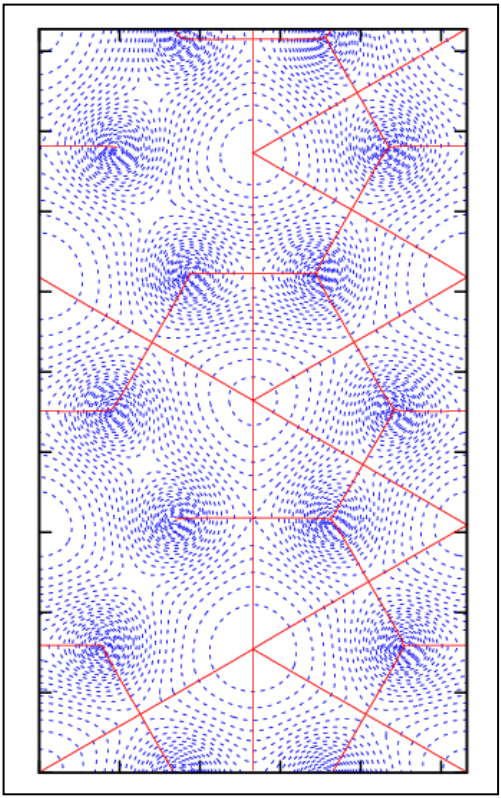

**Fig. 13. The Bader analysis[28] of the G-layers of *Pbcn* hydrogen at 250 GPa. The *ab* plane is shown. The (3, -1) Bond Critical Points are at: 0.236, 0.079, 0.250; 0.730, 0.757, 0.251; 0.500, 0.343; 0.250; 0.000, 0.486, 0.250; 0.243, 0.414, 0.249; 0.00, 0.172, 0.250. The electrons in the graphene layer are not delocalized. The bond critical points of between pair of H atoms can be clearly defined. The $H_2$ remain molecular in nature.**

In conclusion, combined experimental and theoretical studies on the high-pressure phases of hydrogen at 250-300 K reveal a number of new phenomena, which were not expected previously (*e.g.*, Refs. [8,11,33]). The optical and vibrational spectroscopy experiments on phase IV up to 320 GPa shows strong anharmonic effects, consistent with a large atomic motions and strong proton tunneling phenomena. The MD simulations document the strong atomic motion for phase IV (which makes the averaged in time structure to be *Ibam*), and in addition demonstrate a spontaneous transition to the *Cmca*-4 structure. Electronic band structure calculations performed on *Pbcn* hydrogen, used as a structural proxy for phase IV, demonstrate the intermolecular charge transfer that opens an indirect band gap in good accord with the experimental observations. Both studied high-pressure phases of hydrogen show the electronic density anomalies, which reduce the intramolecular bond order index. Further experiments to higher pressures are needed to verify the predicted phase changes reported here.

A. F. G. acknowledges support from the NSF, Army Research Office, NAI, and EFRee. V.V.S. acknowledges support of the DOE Grant No. DE-FG02-02ER45955. R.H and E.G. acknowledge support from the U.K. Engineering and Physical Sciences Research Council and Institute of the Shock Physics, Imperial College. We thank C. J. Pickard for important comments on the results concerning the charge density, and for sending us the results of phonon calculations for the *Pc* structure of phase IV of $H_2$.